# Tailored nano-electronics and photonics with two-dimensional materials at terahertz frequencies

Leonardo Viti,[1] Miriam Serena Vitiello[1]

[1]NEST, CNR—Istituto Nanoscienze and Scuola Normale Superiore, Piazza San Silvestro 12, 56127, Pisa, Italy

miriam.vitiello@sns.it

**Abstract.** The discovery of graphene and its fascinating capabilities have triggered an unprecedented interest in inorganic two-dimensional (2D) materials. Van der Waals (vdW) layered materials as graphene, hexagonal boron nitride (hBN), transition metal dichalcogenides (TMDs), and the more recently re-discovered black phosphorus (BP) indeed display an exceptional technological potential for engineering nano-electronic and nano-photonic devices and components "by design", offering a unique platform for developing new devices with a variety of "ad-hoc" properties. In this perspective article, we provide a vision on the key transformative applications of 2D nanomaterials for the developments of nanoelectronic, nanophotonic, optical and plasmonic devices, at terahertz frequencies, highlighting how the rich physical phenomena enabled by their unique band-structure engineering can allow those devices to boost the vibrant field of quantum science and quantum technologies.

## 1. Introduction

The terahertz (THz) region of the electromagnetic spectrum (1-10 THz frequencies, 300-30μm) has historically suffered from a lack of photonic and optoelectronic devices, in large part because of inadequate optical materials with suitable reflectivity, absorption or transmission in this spectral range. However, emerging applications such as quantum sensing [1], high data-rate communications [2], security [3], metrology [4], atmospheric sensing [5], biomedical imaging [6,7], processing and quality control [5], food inspection [8] and cultural heritage [9] are increasing the demand for reliable, miniaturized and cost-effective THz technologies, easy to be integrated with existing photonic or electronic platforms.

A widespread technological development, in this spectral range, entails the development of suitable sources, modulators and detectors. However, it is well known that, while the THz spectral region lies at the boundary between the mature and well-developed domains of infrared photonics and high-speed electronics, device concepts taken from the neighboring fields cannot be adapted to THz frequencies [10,11]. From one side, conventional electronic devices are too slow to allow switching, modulation, or actuation at THz frequencies [12]. From the other side, traditional semiconductor materials - a cornerstone of photonic devices - have a bandgap that far exceeds the THz photon energy, making them unsuitable for devising active optoelectronic components such as detectors, sources and modulators.

Artificial semiconductor heterostructures played a major role in the development of THz electronic and photonic technologies [13,14], providing a highly effective platform for the manipulation and control of carriers and photons, and overcoming the fundamental limitation represented by the valence to conduction band gap. However, these solid-state architectures require challenging epitaxial growth procedures [15] and, commonly, low operation temperature [16,17].

Layered two-dimensional (2D) materials, as graphene [18,19], black phosphorus (BP) [20], transition metal dichalcogenides (TMDs) [21], and their related van der Waals (vdW) heterostructures [22], have garnered an increasing attention in the last few years owing to their unusual optical and electrical properties [23] that can enable manipulation, propagation and detection of THz waves with an incredible level of control.

## 1.1 Graphene

Graphene is an attractive material in optoelectronics, and, because of its vanishing bandgap (Figure 1a) [19], which results in a broadband absorption, has often been regarded as a magic material in nanoelectronics and photonics. Beside the interband absorption, driven by optical transitions between the valence and conduction bands, light-matter interaction in graphene is also characterized by a strong intraband absorption [24], which makes it promising for the detection, generation, and modulation of THz waves.

Interestingly, the absorption of THz radiation is accompanied by the heat transfer towards the electronic sub-system. The record-low electronic specific heat ($\sim 2000$ $k_B \mu m^{-2}$ at 300 K, where $k_B$ is the Boltzmann constant [25]), stemming from the light-like linear energy dispersion, entails ultrafast carrier thermodynamics [26,27] that results in large thermal gradients and huge non-linear response to THz transients [28]. Coupled with high room temperature (RT) carrier mobilities ($\sim 10^5$ $cm^2V^{-1}s^{-1}$ [29-31]) and high breakdown current density ($\sim 10^8$ $Acm^{-2}$) [32], the ultrafast photoresponse enables high-speed device operation. These unique properties, combined with the possibility of tuning via electrostatic gating, are driving a remarkable surge of research efforts to merge the physics of graphene with device principles across the THz.

## 1.2 Transition Metal Dichalchogenides.

Among the realm of layered 2D materials, transition metal dichalchogenides (TMDs) stand out for their agile large area growth techniques and for the opportunity to engineer their electrical and optical properties by design, i.e. by selecting their stoichiometry and thickness. Indeed, TMDs are $MX_2$-type compounds, where M is a transition element from groups IV, V, and VI of the periodic table and X represents the chalcogen species S, Se, and Te, displaying an array of electronic properties ranging from semiconductor, semimetal, metal to superconductors, depending on the choice of the transition metal. Because of this, they are currently employed in a plethora of applications, encompassing integrated and logic circuits [33], optoelectronic devices, light generation and harvesting [34], valleytronics [35] and neuroscience [21].

Owing to the tunable by design energy gap [21] and the possibility of electrostatic gating, which is a common feature to layered materials, TMDs probably represent the most promising 2D technology for switching applications [36].

TMDs can be used to devise robust ultra-thin-body field effect transistor (FETs) with on-off ratios up to $\sim 10^8$ [37], modulation speeds up to $\sim 50$ GHz [38] and with drive currents $\sim 500$ $\mu A \mu m^{-1}$ on flexible substrates [39], however, with a much lower mobility with respect to graphene ($\sim 10$-100 $cm^2V^{-1}s^{-1}$ [39]). Furthermore, having band gaps ranging from 1 to 2.5eV, TMDs show strong interaction with light from the ultraviolet (UV) to the near-infrared (nIR) frequency ranges, as a consequence of excitonic and interband transitions, enabling applications that well complement graphene capabilities. For example, the efficient production of extra-free-carriers under high-energy light illumination has been successfully used to devise high-performance all-optical THz modulators [40,41].

## 1.3 Black Phosphorus.

Black phosphorus (BP), the most stable allotrope of the phosphorus element, is a single-element van der Waals semiconductor material with a honeycomb structure, whose thickness-dependent direct band gap spans from 0.3 eV (bulk) to 1.0 eV (monolayer), bridging the gap between the zero-gap graphene and the relatively wide band gap of TMDs [42]. As such, it presents an intermediate mobility (~1000 cm$^2$V$^{-1}$s$^{-1}$ at RT) and intermediate switching properties (on-off current ratios up to ~10$^5$ in FET configuration) [43], between graphene and TMDs. Interestingly, unlike layered crystals with flat in-plane lattice, the hexagonally distributed phosphorus atoms are arranged in a puckered structure rather than in a planar one, defining two distinguishable in-plane directions, with different electronic band structure, which results in large electrical, thermal and visible/near IR-optical in-plane anisotropy [20]. Thanks to this property, BP has emerged as a fascinating and versatile material for high-frequency electronic and photonic applications, being successfully employed as a saturable absorber in nIR and visible fiber lasers, as a photoconductive switch [44] and as a sensitive and ultrafast photodetector in the telecom band [45]. In the THz frequency range, it has been employed as a broadband room temperature detector [46,47]. However, in spite of these promising application prospects and the huge research efforts in devising novel BP-based architectures, a further technological up-scaling of this material system has been substantially hindered by the lower degree of technological maturity if compared to graphene and TMDs. The reason for this is twofold. First, BP suffers from a strong degradation under ambient conditions as a consequence of the surface interaction with atmospheric H$_2$O vapor or reactive oxygen species [48,49]. This issue can be mitigated via dielectric or hBN encapsulation, or protection with oxygen-sequestering ionic liquids. However, these strategies are both complicated and time consuming. Second, the fabrication of few-layer BP flakes has been typically limited to the top-down approach of micromechanical exfoliation, whereas the controllable bottom-up large-scale growth of few-layer BP films has been a long-standing obstacle. In this regard, very recently, Wu et al. [50] demonstrated the possibility of growing ultrathin BP on the centimeter scale through pulsed laser deposition (Figure 1b), opening the way for further developing BP-based wafer-scale devices with potential applications in nano-optoelectronics and THz photonics.

## 1.4 Van der Waals Heterostructures.

Recent advances in the layer-by-layer assembly of 2D materials has led to the realization of planar vdW heterostructures [22] where the properties of stacked layered materials can be engineered to reach novel functionalities, e.g. record high mobility in hBN-encapsulated graphene [30], anomalous superconductivity in twisted bilayer graphene [51], tailored hot-electron cooling dynamics [52], improved chemical stability [53]. Importantly, 2D materials can be also easily integrated with specific optoelectronic architectures, such as complementary metal-oxide semiconductor (CMOS) circuitry, photonic waveguides, fiber lasers, metamaterials (MMs), thereby expanding the range of performances of existing components. The maturity of large-area growth and delamination/transfer methods [54] and the compatibility with silicon photonics and CMOS platforms are paving the way towards the industrial up-scale of 2D materials-based technologies, which is also proved by the substantial increase of the number of companies participating in this technology space [23].

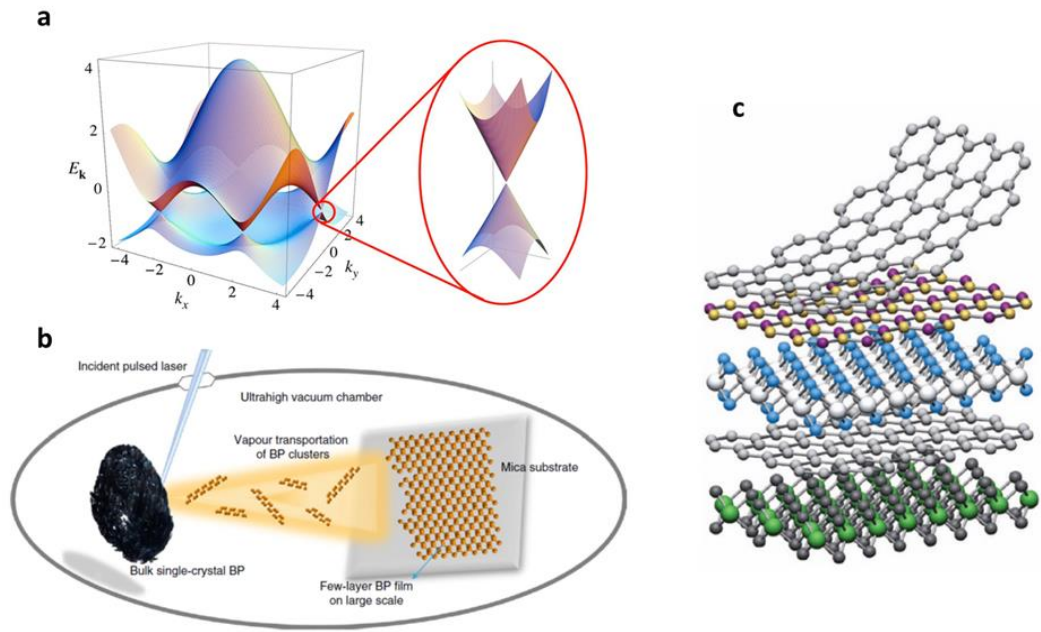

**Figure 1.** (a) Electronic dispersion of graphene. The conduction band and the valence band cross at six discrete points. These points are called *K* points. The zoom shows that the dispersion relation close to the *K* points looks like the energy spectrum of massless Dirac particles [19]. (b) Schematic of the recently developed growth of centimetre-scale few-layer BP films by pulsed laser deposition [50]. (c) Isolated atomic planes of 2D materials can be assembled into designed van der Waals heterostructures, made layer-by-layer, in a precisely chosen sequence [22]. Figure (a) is reprinted with permission from A.H. Castro Neto *et al.* Rev. Mod. Phys. 81, 109 (2009). Copyright 2009, by the American Physical Society. Figure (b) is reprinted with permission from Z. Wu *et al.* Nat. Mater. 20, 1203 (2021). Copyright 2021, Springer Nature. Figure (b) is reprinted with permission from A.K. Geim *et al.* Nature 499, 419 (2013). Copyright 2013, Springer Nature.

In the following we provide a perspective on key device technologies that, taking advantage of the unique properties of 2D nanomaterials, promise groundbreaking impacts in THz photonics and optoelectronics.

## 2. Terahertz detectors

The ability to convert light into an electrical signal, with large quantum efficiencies and controllable physical dynamics, is a major need in photonics and optoelectronics. THz frequency detectors have been the forefront of an intense interdisciplinary research in the last decade, encompassing the investigation of different technologies between optics and microwave electronics, different physical mechanisms and a large variety of material systems [55]. In this context, the rich physics of 2D materials, their high carrier mobility [30], ultrafast light-matter interaction [56], tunable optoelectronic response and room temperature operation, can open new paths towards the realization of novel enabling technologies.

Photodetectors operating in the mid-infrared and THz are key elements for a broad range of applications, spanning from biomedicine [57], security [3,6], spectroscopy [58], cultural heritage [59], environmental monitoring [5], astronomy [60], real-time imaging [61], wearable devices and high data-rate communications [2]. Correspondingly, the state-of-the-art currently includes ultrasensitive technologies operating at cryogenic temperatures, such as superconductive hot electron bolometers [62], fast and sensitive single pixel detectors, such as Schottky diodes [63] or high-electron mobility transistors (HEMTs) [64], and portable THz cameras based on FETs or microbolometer arrays. The sensitivity of THz detectors is usually expressed in terms of noise equivalent power (NEP), defined as the ratio between the noise figure and the device responsivity (amplitude of the electrical output per unit input optical power). A detailed enumeration of all the available devices and related performances can be found in recent topical reviews [55,65]. In the following, we discuss specific applications in which 2D materials can play a disruptive role with respect to alternative material systems.

Photodetection of THz light in 2D materials [66] can be accomplished by several different mechanisms like photo-thermoelectric (PTE) [67,68], ballistic rectification [69], photo-bolometric (PB) [70], plasma-wave (PW) rectification [71] or via a combination of them [66,72]. Importantly, in 2D material based architectures, the dominant detection mechanism can be tailored by design [66]. Their ultrafast dynamics [26] and the ease of fabrication [23,54] and integration [73], can allow boosting both the sensitivity and speed of THz photodetectors. Single layer graphene (SLG), BP and TMDs have been used to fabricate a variety of efficient THz detectors, relying on PB, PTE or PW effects [67-71, 74-81].

## 2.1 Graphene

The peculiar carrier thermodynamics in graphene [52] is ideal for thermally driven photodetection, in particular for the full exploitation of the bolometric and thermoelectric effects [66]. Indeed, when the incoming THz radiation interacts with a graphene layer, the electromagnetic intensity is transferred into internal energy of the electron gas by Drude (intraband) absorption. The electronic subsystem shows a record-low specific heat, which can lead to the ultrafast (50 fs) onset of thermal gradients [27,82] and to a rapid overheating of the electronic distribution with respect to the graphene lattice. This is due to the difference between the electron-electron scattering time ($\sim$ 20 fs [26,83], needed for thermalization of the electronic distribution) and the slower ($\sim$2 ps [84-87]) electron-phonon relaxation time. Therefore, high-energy carriers remain thermally decoupled from the crystal lattice and a *quasi*-equilibrium state is reached, where the electronic temperature, $T_e$, is considerably higher than the lattice temperature $T_L$ [88]. In the case of bolometers, the cooling dynamics allows detection speed with timescales of few ps [70,89]. However, the relatively weak bolometric coefficient of graphene ($dR/dT$ <1%/K [90]) limits the sensitivity of devices based on the readout of the electrical resistance. Interestingly, this can be circumvented by measuring the Johnson-Nyquist noise, or the switching currents in low-temperature graphene based Josephson junctions. In the case of PTE-driven rectification, the efficient carrier heating is accompanied, in high-quality graphene (mobility $\sim 2\times10^4$ $cm^2V^{-1}s^{-1}$), by a large Seebeck coefficient (>150 $\mu VK^{-1}$) [91], which provides a strong thermoelectric conversion. Differently from the bolometric detection, which relies on the THz-induced conductivity change in the material as a consequence of heating, the PTE mechanism requires a certain degree of asymmetry in the graphene device to sustain the thermal gradient that drives the photocurrent. In order to comply with this requirement, different architectures have been employed, making use of asymmetric contacts [67], asymmetric antenna coupling [80], or asymmetric *p-n* junctions [68,81]. The latter takes advantage of the Fermi level ($E_F$) tunability in the graphene layer by electrostatic gating, which results in the possibility of changing the sign of the Seebeck coefficient on the two sides of the junction (Figure 2a), therefore allowing the optimization of detector responsivity by selecting the proper gate voltage configuration. Importantly, the non-monotonic dependence of the Seebeck coefficient from $E_F$ leads to multiple sign changes in the photocurrent when the carrier density in either side of the junction is tuned, giving rise to the so-called six-fold pattern (Figure 2b), a distinctive feature of the PTE effect [92]. Taking advantage of the unique graphene properties, PTE graphene-based THz detectors have achieved a superior combination of performances (sensitivity, dynamic range, speed) [67,68,80,81] at frequencies above 2 THz with respect to commercially available technologies (pyroelectric detectors, Golay cells, Si-bolometers, microbolometers, MOSFETs) [55]. Recently, antenna-coupled PTE *p-n* junctions with NEP ~80 pWHz$^{-\frac{1}{2}}$ and sub-ns response times [68,81], operating at 3 THz and at RT, have been demonstrated in high-quality hBN encapsulated graphene. In parallel, a recent study on microwave bolometers [93] based on superconductor-graphene-superconductor (S-g-S) operating at T<1K demonstrated performance levels at the threshold for circuit quantum electrodynamics, with unprecedented NEP levels ~30 zWHz$^{-\frac{1}{2}}$. Therefore, hot-carrier-driven THz photodetectors in graphene present superior performances with respect to other commercially available long-wavelength technologies.

Beside PB and PTE effects, another detection mechanism which plays an important role in graphene-based field effect transistors (GFETs) is the PW or Dyakonov-Shur effect [94]. This effect enables the rectification of an *ac* field at frequencies higher than the intrinsic transit time-limited cutoff of the transistor. In their room-temperature implementations, PW-driven FET THz detectors operate in a nonlinear second-order regime, through overdamped plasma instabilities induced along the FET channel when the THz field, coupled to the source and gate electrodes, simultaneously modulates the carrier density and their drift velocity [66,94]. The resulting current exhibits a *dc* component whose magnitude is proportional to the square amplitude of the *ac* field, hence to the intensity of the incoming radiation, and can be measured at the drain contact either in photocurrent mode or photovoltage mode. Similarly to the PTE mechanism, this effect requires an asymmetry along the channel to occur, thus it often interplays with the PTE to the generation of the photoresponse [71,78]. Interestingly, the PW mechanism has been recently reported in its long-sought resonant form in asymmetric antenna coupled bilayer graphene (Figure 2c) [79]: at sufficiently low temperatures (T<10K) the GFET channel can act as a plasmonic Fabry-Perot cavity endowed with a rectifying element [79] (Figure 2d). The photoresponse then shows distinct peaks as a function of the gate bias, whose positions carry information on the plasmon wavelength (Figure 2e). Owing to this feature, this technology has been proposed as a helicity-sensitive plasmonic interferometer, opening up potential new applications in on-chip polarization and phase-shift analysis of coherent THz beams [95].

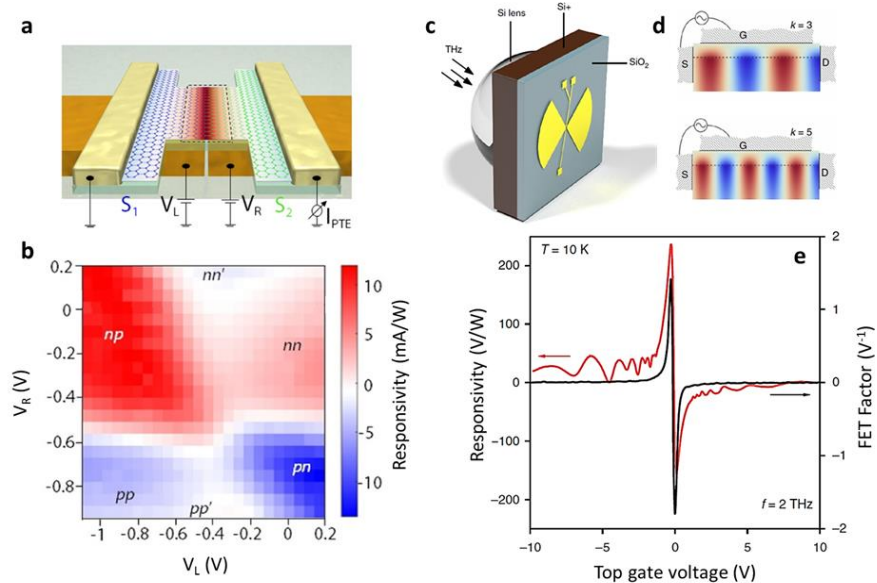

**Figure 2.** Photo-thermoelectric and plasma-wave mechanisms in graphene FETs. (a) Schematic representation of an antenna-integrated *p-n* junction THz detector, where the gate voltages applied on the two sides ($V_L$ and $V_R$) create a difference in the Seebeck coefficient (*S*) across the junction. When the device is heated by an incoming THz beam, the electromagnetic energy is funneled at the center of the junction (red-shaded area) by the antenna and a photo-thermoelectric current ($I_{PTE}$) is generated. (b) Responsivity as a function of voltages applied to the two antenna branches/gates; the six-fold change of polarity is a distinctive characteristic of the PTE response [68]. (c) Bow-tie antenna coupled GFET, mounted on a silicon lens for better optical coupling. (d) Plasmon resonances in the gated channel of an encapsulated graphene FET. In the resonant PW mechanism, the FET channel acts as a plasmonic Fabry-Perot cavity. (e) Responsivity of a GFET in the resonant plasmonic regime (red curve). The black curve shows the trend of the photoresponse expected for the broadband overdamped PW mechanism, which is proportional to the FET factor [79]. Figures (a) and (b) are adapted with permission from S. Castilla *et al.* Nano

## 2.2 Black Phosphorus

If graphene is an interesting material system for applications in the infrared, due to its high mobility and gapless nature, the inherent in-plane anisotropy of BP, combined with its tunable band gap and high Seebeck coefficient (up to ~0.4mVK$^{-1}$ at 300 K [96]), makes it an appealing and intriguing alternative to reduce the large dark currents that dominate in the gapless graphene, under non-zero bias operation. The $sp^3$ hybridization of P atoms arranged in a 2D honeycomb lattice leads to the formation of a puckered structure as a consequence of electrostatic repulsion [97]. The puckering is periodic along the armchair direction ([*001*], *c*-axis) and orthogonal to the zigzag direction ([*100*], *a*-axis). The in-plane crystallographic anisotropy affects the band structure, in turn inducing the anisotropy of its optical (linear dichroism [98]), electrical (mobility, effective mass [99]) and thermal (thermal conductance [100]) properties, opening the possibility to selectively activate different THz detection dynamics in a BP-FET, by choosing the relative orientation between the armchair axis and the THz antenna axis (polarization axis) [46,101]. Taking advantage of this peculiarity, different detector geometries have been devised, leading to RT PTE detectors with NEPs~5nWHz$^{-\frac{1}{2}}$ at 3.5 THz and ~μs response times [77] and NEPs~58pWHz$^{-\frac{1}{2}}$ at 0.1 THz, in an ultrashort channel BP-FET [102], and with signal to noise ratios up to 20000 when embedded in a vdW heterostructure [47].

## 2.3 Challenges and opportunities

### 2.3.1 Imaging

THz rays have an essential role in many imaging applications, from industrial inspections to medical diagnosis. However, commercialization is prevented by impractical and expensive available THz instrumentation.

The first demonstration of THz imaging with a graphene based device [71] has induced an exciting burst of research activities. In this regard, zero-bias, zero-power consumption PTE detectors represent an undisputed advantage to meet the requirements of low cost and low SWaP (size, weight and power) of room-temperature THz cameras. However, the aforementioned competitive performances in graphene based PTE detectors have been so far realized in single-pixel architectures, which still limit the technological up-scaling of the proposed devices and their application to THz imaging systems. The current challenges in this field, beside the improvement of performances, are mainly related to the development of reliable integration techniques and to the implementation of a technological flow chart fully compatible with standard CMOS readout integrated circuits. Thanks to the recent advances in the large-scale growth and assembly of graphene, hexagonal boron nitride (hBN) [103] and black phosphorus [50] thin films, we envision a near-future impulse of 2D material-based devices towards industrial-scale, multi-pixel THz cameras.

### 2.3.2 Communication

Wireless networks are evolving to respond to the ever-growing request for higher data rates. The 5G scenario is characterized by impressive services in terms of system. However, it is well known that the mobile data traffic has been exponentially increasing for more than a decade and this trend is expected to continue for the foreseeable future [104,105]. The fast growth of the Internet of Things (IoT) and the emergence of new services (e.g. extended reality services, telemedicine, haptics, flying vehicles, brain-computer interfaces, and autonomous systems) require ultra-high capacity (1Tbps) connectivity to support high density (1000× with respect to 5G) access networks with reliable and low end-to-end delays (<1 ms) [106]. Therefore, the 5G systems gigabits per second rates may fall short in supporting many emerging applications that require several hundreds of gigabits per second to several terabits per second data-rates with low latency and high reliability, which are expected to be the design goals of the next generation 6G communications systems [107]. THz (and millimeter) waves will provide new means and radically new technologies for wireless communication, at speeds that by far overcome today's cellular and WiFi networks [108,109]. Given the potential of THz communications systems to provide the required data-rates over short distances, they are widely regarded to be the next frontier for wireless communications research. With the recent advancements in device performances, THz communication is expected to play a pivotal role in the upcoming generations of communication standards [110]. In this regard, graphene based PTE detectors, with expected speed limit >100 GHz, low energy consumption, low footprint and weight requirements, broadband and room temperature operation, and CMOS compatibility, intrinsically ensure the possibility of large volume manufacturing and low-cost, thus matching the future requirements for the 6G era.

### 2.3.3 Quantum Information

Quantum information science is rapidly progressing and, although it is hard to predict how the field will develop in the future, it seems clear that photonic qubits (flying qubits), thanks to their intrinsic mobility and stability in time, will play an important role in quantum communication. One of the most important building blocks for linear optical quantum computing (LOQC) is the single-photon detector, which simultaneously represents one of the major challenges in the far-infrared domain. To date, THz photon counting has been only realized in two platforms: quantum dots capacitively coupled to single electron transistors (T= 50 mK, needs magnetic field) [111] and superconductor-based quantum capacitance detectors (T= 15mK) [60]. Both technologies require deep cryogenic cooling and their wide implementation seems unfeasible. In this direction, new types of detectors based on 2D materials are emerging as promising ultrasensitive alternatives. Notable examples are the aforementioned S-g-S architecture with NEP ~30 zWHz$^{-\frac{1}{2}}$ [93], magic-angle bilayer graphene nanocalorimeters [112], capable of detecting single photons of ultralow energies by taking advantage of the record-low heat capacity and sharp superconducting transition, tunnel FETs based on bilayer graphene [113], negative capacitance FETs based on $MoS_2$ [114] and graphene quantum dots devices [115]. Our guess is that, in the near-future, a ground-breaking quantum technology based on a 2D platform is likely to emerge from this burst of scientific effort.

## 3. Modulators

Optical modulators can allow manipulating the amplitude, phase, frequency or polarization of a radiation source [116]. With the recent evolution of wired and wireless communication networks and the exponentially increasing demand for high data-rates and services [105], modulators in the nIR and visible frequency ranges have matured significantly enabling terabit-data communications by means of all-optical devices and circuits [117,118]. On the other hand, THz technology has not yet reached the same maturity level, required to be commercially viable in the near-future. In this regard, owing to their versatility, integrability [119] and unusual and tunable optical properties, 2D materials can be a game-changer in the quest for a breakthrough in the realm of THz modulators.

Optoelectronic modulators are widely used in modern telecommunications [120]. As wireless communication is approaching the THz range [108,121], the development of efficient modulators is becoming progressively more urgent. In the last decade, different approaches, based on III-V semiconductors such as silicon and gallium arsenide, or employing 2DEG in AlGaAs/InGaAs heterostructures were demonstrated, with modulation speeds up to 14 GHz and room temperature operation [122,123]. However, the high demand for fast, efficient, integrated amplitude, frequency, and polarization modulators operating at room temperature is driving extensive research on 2D material-based devices that, being easy to be combined with other material systems and devices, can allow disruptive large-scale applications [124]. On the other hand, their small interaction volume severely limits the modulation contrast, orienting present research efforts toward metamaterial-based architectures and/or waveguide coupling. In the following, we discuss the most relevant modulator architectures.

## 3.1 Amplitude modulators

### 3.1.1 Broadband amplitude modulators

An amplitude modulator is conventionally defined by means of two main figures of merit [72]: the modulation depth $m$ (or contrast), defined by the percent change in transmission or reflection (in %) and the cut-off modulation frequency, i.e. the frequency that reduces $m$ by 3dB. Typically, amplitude modulation is achieved by changing the optical absorption of a material. In the THz range, the optical properties of 2D materials are dominated by free-carrier (Drude) intraband absorption, which is proportional to the real part of the complex conductivity [56]. Therefore, changing the free-carrier density $n$ or the free-carrier temperature $T_e$ results in a change in the absorption [72]. The most technologically relevant means of controlling the THz optical properties of a 2D material are the optical or electrostatic gating of the carrier density and carrier temperature. In recent years, a surge of research activity has boosted major advances in this field, heading towards two main directions: integrated all-optical and all-electronic modulators (Figure 3a,b) [125].

THz optically-driven intensity modulation relies on the modification of a device property (e.g. transmission), obtained by inducing a change in the material electronic system by illuminating it with a pump field. Depending on the energy of the pump photons and on the initial doping of the material, the carrier density $n$ or the electronic temperature $T_e$ (or both) can be modified, resulting in a visible change of conductivity.

All-optical modulation has historically represented the most convenient and technologically relevant solution in the THz frequency range, thanks to the opportunity of utilizing silicon [126] or III-V semiconductors [127] such as germanium or GaAs as active substrates, together with near-infrared pumping light. This scheme is intrinsically broadband and presents a high degree of versatility, being able, for example, to create a custom spatial distribution of free carriers in a substrate that, in turn, can act as a programmable spatial THz light modulator [128]. Both the modulation depth and speed are dictated by material properties: the carrier mobility and conductivity indeed limit the modulation depth, whereas the recombination time ($\tau_r$) of photo-generated carriers sets the limit for the speed ($\tau_r > 1$ ms for silicon [129] and $\tau_r < 1$ ps for low-temperature grown GaAs [130]).

In this context, combining 2D materials with semiconductor wafers can increase the all-optical modulator performances. For example, the integration of monolayer graphene with silicon has led to modulation depth up to the 80% (Figure 3a) [131], and the combination with germanium can increase $m$ from the 64% of the substrate alone up to the 94% by pumping at 1550nm [132]. Such an improvement is related to the transfer of photo-generated carriers from the substrate to the graphene layer, where they experience a higher mobility, and in turn, a higher conductivity. Similar results were obtained by combining TMDs ($WSe_2$, $MoS_2$) with semiconductor wafers [133]. Interestingly, integrating a 2D material with a semiconductor substrate allows selecting by design the timescales governing the carrier dynamics, depending on the carrier cooling channels available in the 2D materials. More importantly, layered materials bring the additional opportunity of tuning the optical modulation by means of an external voltage between the 2D layer and the underlying substrate [134], opening a novel route for the development of hybrid optoelectronic active modulators.

The alternative domain of all-electronic amplitude modulators has been pioneered by extensive works on large-area graphene by Maeng and Sensale-Rodriguez in 2012 [135,136]. Since then, it became clear that, thanks to its unique tunable optical properties and to the ease of integration with existing photonic and electronic platforms, large-area graphene could become a game-changer in all-electronic THz modulators. Tunable and broadband amplitude modulation can be simply achieved by applying a proper gate voltage, capable of adjusting the Fermi level in the graphene sheet. Different geometries and architectures have been proposed operating both in transmission and reflection mode, with modulation depths beyond 60% and modulation speed in the 100 kHz range. The use of ionic-liquid gating can boost $m$ up to the 99% [137] (Figure 3c,d), also enabling the realization of flexible devices [138], but limiting, on the other hand, the modulation speed.

### 3.1.2 Resonant amplitude modulators: hybrid systems and metamaterials

A convenient way to tackle the issues of low absorption, which limits $m$, and low modulation speed is represented by the combination of 2D materials with planar metallic structures, such as antennas [139] or metamaterials [140] (Figure 3b). These architectures are widely used in optics and their use have been extended in the last decade to the THz domain [141]. In particular, metasurfaces offer unprecedented functionalities for beam shaping, polarization control and wavefront generation [142]. Thanks to the broadband and tunable absorption of graphene, MMs are an ideal platform to create structures with tailored properties in the far-infrared. The integration of a lossy material within a metasurface can exploit the field-enhancement induced by the metallic elements to increase the absorption in the 2D layer. A direct consequence of the presence of a metamaterial is the reduction of the overall THz bandwidth, which is fixed by the geometry and dimensions of the individual MM elements [123,143]. Tuning the Fermi level of an MM-coupled graphene layer results both in a change of the intraband absorption [72] and in a shift of the MM resonance driven by the change of the real part of graphene's conductivity. This combined effect is typically stronger than the broadband modulation of the absorption alone, and, most importantly, enables the simultaneous control on the amplitude and phase degrees of freedom of an electromagnetic wave. Interestingly, integrating a 2D material with a metallic metasurface increases the *local* interaction with THz radiation, allowing the reduction of the absorbing area in correspondence to the positions of maximum near-field enhancement. This strategy has been successfully used in graphene based split-ring geometries to reduce the graphene footprint by two orders of magnitude [144], therefore pushing the modulation speed in the MHz range, and in plasmonic bow-tie antenna arrays coupled to graphene ribbons to go even beyond 100 MHz [139], and has the potential to improve the cutoff frequency up to the GHz range with further parameters optimization [145].

Interestingly, the metasurface coupling is applicable to any large area 2D material and to all-optical modulators as well. Notable performances have been obtained in $MoS_2$-Si hybrid systems, with transmittance modulation up to 90% [146] and in an all-optical device based on THz asymmetric metamaterial arrays (TASR) integrated with bulk $MoS_2$, with an estimated switching time of 100 ps [147] and even in THz switches based on inorganic perovskite quantum dots [148].

An electrically switchable graphene THz amplitude modulator with a tunable-by-design optical bandwidth has also been recently demonstrated [149]. It comprises a grating-gated graphene capacitor on a polyimide quarter wave resonant cavity. The core principle is the electrostatic gating of a single layer graphene, achieved by a metal grating used as a gate electrode, with an $HfO_2/AlO_x$ gate dielectric on top. This is patterned on a polyimide layer, which acts as a quarter-wave resonance cavity, coupled with an Au reflector underneath; 90% modulation depth of the intensity, combined with a 20 kHz electrical bandwidth in the 1.9–2.7 THz range has been demonstrated [149].

### 3.2 Polarization Modulators

A compelling application for hybrid systems involving 2D materials metasurfaces is the direct manipulation of the polarization state of an incoming THz beam. This is of paramount importance for THz communications, enabling protocols such as polarization shift keying [150] and polarization division multiplexing [151]. Conventional polarization modulators in the THz range are based on micro-electromechanical systems (MEMS) or photoactive materials combined with 3D chiral structures [125], but both concepts present intrinsic limits: MEMS have low maximum modulation speed, whereas photoactive schemes require complex optical systems.

Interestingly, the combination of tunable graphene with chiral planar metasurfaces can strongly simplify the architecture: an all-electrical polarization controller has been recently employed to selectively modify the polarization state of a THz QCL, with potential modulation speed in the 100 MHz range [152]. Furthermore, the electrostatic control of magnetic circular dichroism [153] and Faraday rotation [154] can also be achieved in graphene-coupled magnetoplasmonic metasurfaces (Figure 3e), however, at the cost of an applied external magnetic field.

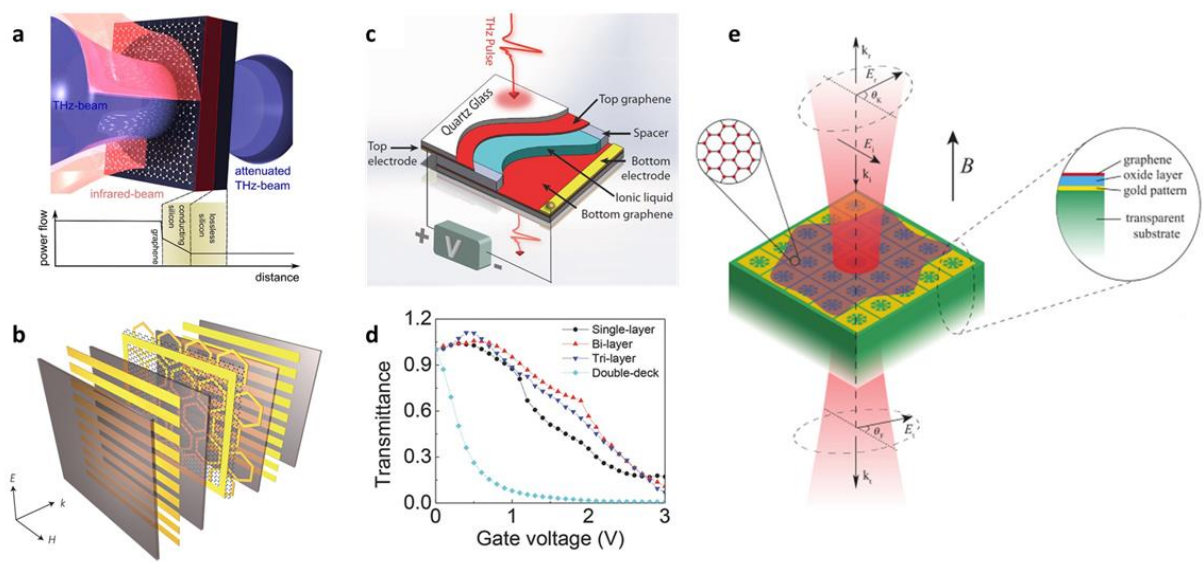

**Figure 3.** Graphene-based THz modulators. (a) Schematic view of an all-optical terahertz wave modulator [131]: a near infrared beam induces photodoping in graphene, leading to a modulation depth of 99%. (b) Gate controlled active graphene metamaterial, composed of a single layer graphene deposited on a layer of hexagonal metallic meta-atoms [140]. (c) Structure of an ionic-liquid based graphene modulator [137]. When a gate voltage is applied between the top and bottom electrodes, it induces charges on the graphene layers, altering their transmittance, as shown in (d), for different numbers of graphene layers. (e) A magneto-optical THz structure composed by a monolayer graphene and a metallic metasurface of optical resonators [154]. Figure (a) is reprinted with permission from P. Weis *et al.* ACS Nano 6, 9118 (2012). Copyright 2012, American Chemical Society. Figure (b) is reprinted with permission from S.H. Lee *et al.* Nat. Mater. 11, 936 (2012). Copyright 2012, Springer Nature. Figures (c) and (d) are reprinted with permission from Y. Wu *et al.* Adv. Mater. 27, 1874 (2015). Copyright 2015, John Wiley and Sons. Figure (e) is reprinted with permission from A. Ottomaniello *et al*. Optics Express 26, 3328 (2018). Copyright 2018, The Optical Society.

### 3.3 Phase Modulators

The change in the complex conductivity of 2D layered materials when biased by an electric field allows controlling the phase of the beam that is reflected or transmitted by the modulator. However, broadband Drude absorption typically has a minor effect on the phase of the output THz radiation [155]. The combination of a thin lossy material (e.g. graphene) with a cavity [156] or a resonant metasurface [157] can significantly boost this effect. In the latter case, this is due to an induced change in the resonance frequency of the metamaterial, rather than in the broadband absorption: a variation in the carrier density, e.g. induced by electrostatic gating, translates in a spectral detuning of the resonator response, which can result in huge phase changes up to 180° [157]. Interestingly, metasurfaces are natural arrays, where each unit cell can be individually addressed with a different electrostatic tuning in order to achieve arbitrary phase landscapes. Therefore, the integration of passive metamaterials, composed of prearranged artificial unit cells possessing different electromagnetic responses, with atomically thin, tunable materials has unleashed the development of coding and programmable metasurfaces for digital THz flat optics [158]. To date, the frontier is represented by reconfigurable reflect-arrays based on monolayer graphene capable of achieving beam steering, shaping and broadband phase modulation [159]. Interestingly, graphene reconfigurability can be exploited in the realization of programmable THz sensors based on plasmonic antennas [160], or programmable THz radiators based on leaky wave antennas [161]. Related applications are summarized in a recent review [162].

### 3.4 Frequency Modulators

The most commonly employed strategies to devise a frequency modulator rely on tuning the lumped components of a resonant circuit [141,143,163] or inducing a change in the plasmonic resonances in a material [164-167]. Complementary split ring resonator (C-SRR) arrays strongly coupled to graphene ribbons have been proposed as a valuable solution for efficient frequency modulation [167]. In these architectures, tunable plasmonic graphene resonators are coupled to conventional split-ring resonators, by fabricating graphene ribbons at the center of the C-SRR structure. The frequency response of this hybrid system can be efficiently modified acting on the carrier density in the graphene layer by means of electrostatic gating, leading to a 60% modulation depth with a 3dB cutoff frequency of ~41 MHz [167].

### 3.5 Integrated systems: waveguides and quantum cascade lasers (QCLs)

Improving the coupling of THz radiation and atomically thin materials via metasurfaces and resonators has been a quite successful approach, but at the expense of a narrow operation bandwidth. Alternatively, incorporating graphene in or on a waveguide facilitates in-plane coupling of either the confined or the evanescent mode of a propagating THz beam, promising almost 100% modulation depths [168]. The first experimental demonstration of a waveguide-integrated graphene THz modulator has shown a modulation bandwidth of 90% between 0.15 and 0.17 THz [155].

Interestingly, the intrinsic versatility of 2D materials can be exploited to transfer them directly onto laser waveguides. As a natural consequence, graphene-based modulators have been directly integrated with QCLs [169,170]. Thanks to the direct proximity with the source and to the strong intraband absorption, the emission of the hybrid QCL-graphene architecture is strongly influenced by the graphene doping, enabling the simultaneous tunability of the output power and frequencies [169], with operating speed in the 100 MHz range [170].

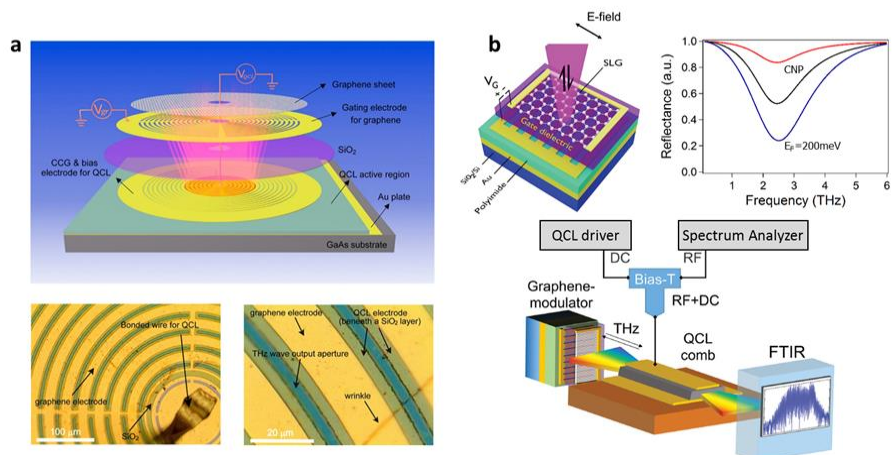

**Figure 4.** Graphene-based modulators integrated with THz QCLs. (a) Schematic of the graphene modulator integrated with a QCL and optical pictures of the real device. Light is emitted vertically from the surface and is modulated by the electrically gated graphene. The inner central rings are connected to the top surface of the QCL for biasing the device, the outer ones are in contact with graphene and are used for modulating the emission [170]. (b) Layout and operating principle of a standalone grating-gated single layer graphene THz modulator. When integrated with a THz QCL frequency comb, the electrically switchable reflectance of the modulator compensates the laser group velocity dispersion, resulting in an improved stability of the comb operation [149]. Figure (a) is adapted with permission from G. Liang *et al.* ACS Photon. 2, 1559 (2015). Copyright 2015, American Chemical Society. Figure (b) is adapted with permission from A. Di Gaspare *et al.* Adv. Funct. Mater. 31, 2008039 (2021). Copyright 2020, John Wiley and Sons.

### 3.6 Integrated systems: Optical resonators for biosensing

Optical biosensors [171] offer great advantages over conventional analytical techniques since they enable the direct, real-time and label-free detection of many biological and chemical substances. The biological information is here retrieved by detecting a change in the optical properties of a transducer, induced by its interaction with the biological sample, the analyte [172], which is typically forced on the surface of the sensing layer.

The THz frequency range is particularly interesting in this respect, because it encompasses the vibrational fingerprints of important macromolecules, like nucleic acids and proteins [173]. Moreover, owing to the recent advances in the micro- and nano-fabrication of designer architectures, the implementation of novel structures and new materials has now become a major research focus in the realm of THz optical sensing. Although optical biosensing at THz frequencies is still in its infancy, if compared to the visible and ultraviolet domains, a number of promising solutions have been recently implemented as optical transducers [174]. As an example, MM-based transducers have been used to quantitatively analyze different concentrations of glucose [175] and viruses [176], taking advantage of the frequency shift in the resonance of an SRR array. Plasmonic metasurfaces have been used to realize optical transducers sensitive to the refractive index of gaseous analytes through the excitation of THz surface plasmon resonances (SPR) [177,178].

Due to their strong physical and chemical interactions at the interface with the environment, 2D materials can offer promising advantages in the development of transducers to be embedded in optical thin-film sensors [179]. In particular, graphene, emerged as an attractive plasmonic material [180], can be potentially appealing to overcome some of the issues related to metallic architectures employed in the THz and mid-IR ranges, specifically the weak field confinement and the limited tunability. Indeed, the SPR in graphene, or graphene/metallic MMs, can be dynamically modulated by electrical gating to sweep over the vibrational modes of the target biomolecule [181], and the electromagnetic fields constituting graphene plasmons (GPs) can be confined down to one-atom-thick layers in graphene-insulator-metal platforms [182]. The extreme light confinement is further accompanied by low losses for the GP modes from the mid-IR [183] to the THz range [184], providing a unique platform to enhance light-matter interaction and optically probe bio-chemical systems with increased sensitivity with respect to the metallic counterparts [185]. In addition to this, graphene chemically interacts with the analyte through delocalized $\pi$-electrons, undergoing dramatic changes in its optical properties as a consequence of the analyte-induced modulation of the chemical potential *i.e.* doping. Thanks to the good biocompatibility of graphene [186], graphene-coated THz MMs have been proposed and successfully employed as aptasensors [187] and in the label-free detection of DNA hybridization [188,189], demonstrating that the chemical coupling between the target DNA sample and graphene significantly improves the detection sensitivity with respect to the bare metamaterial [187].

### 3.7 Challenges and opportunities

### 3.7.1 Graphene plasmonic resonators
Surface plasmons in doped graphene are different from plasmons in noble metals as they can be tuned by gating or doping [190,191]. In the far-infrared, GPs exhibit extreme field-confinement ($<10^{-4}$ $\lambda_0$ [182], where $\lambda_0$ is the free space wavelength), in contrast with the weaker field compression that instead occurs in noble-metal structures. This, however, comes at the expenses of broader resonances in all-graphene metasurfaces. With respect to the metal structures, graphene offers a more favorable trade-off between bandwidth, losses and field confinement, at THz frequencies [192] (below the optical phonon branch $\hbar\omega \approx 0.2\text{eV}$ [193]), that could be conveniently exploited in applications where a strong light-matter interaction is beneficial, e.g. biosensing [185], quantum plasmonics [194] and nano-tweezers [195]. On the other hand, atomically thin graphene has high intrinsic loss and a relatively inefficient interaction with light that severely hamper the achievement of high-quality resonances in all-graphene metamaterials and metasurfaces [196]. Moreover, the required balance between modulation amplitude and losses in graphene resonators poses fundamental limitations on its use as a standalone plasmonic element [197-199]. For this reason, graphene is typically combined with [200], or embedded into [159], MMs based on noble-metal nanostructures. This strategy allows taking advantage of the strong and narrow resonances of the metallic patterns and simultaneously of the graphene tunability, which functions as a lumped element for the metallic resonators: modulating the conductivity of graphene influences the optical response of plasmonic arrays, leading to dynamically controllable active plasmonics. A promising alternative strategy, which entails the use of all-graphene resonators, is represented by the exploitation of loss-compensation mechanisms [201], attainable by carefully tailoring the combination of sufficiently high doping [193] and long relaxation times (high mobility), distributing carriers into multiple graphene layers [202], or exploiting the graphene negative dynamic conductivity [203]. Importantly, long plasmon lifetimes are increasingly difficult to realize for highly confined modes. The recent improvements in large-area graphene quality [204] and the use of van der Waals heterostructures [183,205] are expected to significantly increase the plasmonic lifetime [206], with beneficial outcomes in terms of device performance, such as bandwidth, phase-shift, sensitivity, and modulation efficiency.

### 3.7.2 Programmable metasurfaces
The future of THz modulators certainly goes in the direction of integrated devices compatible with (or realized in) solid-state semiconductor technology (III–V and silicon based) [207] that can operate at room temperature and can be manufactured at a low cost, exploiting economies of scale. 2D materials are therefore an ideal choice, in this respect, since they can offer a versatile and programmable architecture. Programmability could include electronic reconfigurability of the wavefront and polarization of the emitted THz fields, for applications in communication, radar and imaging [208], or dynamic spectral control of the radiated fields for spectroscopy and hyperspectral imaging [209]. Importantly, the broadband nature and ease of integration of graphene offer the possibility to design architectures across the whole THz domain and beyond, in contrast with respect to other material platforms like VO$_2$ [210] or liquid crystals [211,212]. Such versatility is often required for advanced applications, and is generally lacking in many current non-integrated THz platforms.

Programmable active metasurfaces offer multifunctional choices, being able to implement by design different THz modulation patterns, thus enabling the coexistence of advanced performance and high integration. A promising route to fully exploit this advantage is represented by high-sensitivity single-pixel real-time imaging realized via compressive sensing using spatial light modulators (SLMs). While the most straightforward way of image acquisition with a single pixel detector is with raster scan techniques, this can result in prohibitively time consuming systems. Compressive imaging employs a SLM that acts as a programmable transmission mask and could overcome this bottleneck providing technologically relevant acquisition rates and potentially enabling real-time imaging [61]. In this regard, active graphene metasurfaces could be ideally suited for the realization of electrically programmable ultrathin SLMs. However, their evolution is still at an early stage and major advances need to be implemented to increase the current operation speed (100 MHz), bridging the gap with the state-of-the-art, represented by silicon based electro-optic modulators with transmission rates approaching 1 Gbit per second [213] and 2DEG realized in AlGaAs/InGaAs heterostructures that can reach operation speed > 10 GHz at room temperature [122].

Eventually, the integration of graphene programmable metasurfaces with THz QCLs can open new scenarios for all-solid state quantum information processing and quantum metrology. This stems from the possibility to alter the emission state of a multimode QCL with external modulators [149], driving it into the frequency comb regime, where the output lasing modes are perfectly equally spaced and possess a well-defined phase relationship between each other. Thanks to their intrinsic coherence in the frequency and time domains, frequency combs are widely used for ultra-high-resolution spectroscopy, hyperspectral imaging [214], time-domain nanoimaging, quantum science and technology, metrology, nonlinear optics, frequency multiplexing and for the generation of short optical pulses [215]. All these applications are typically addressed by using different architectures and different QCL designs. Programmable metasurfaces could impact this field by introducing a reconfigurable component that can be custom-tailored according to the specific task. In this regard, the versatility and flexibility of 2D materials can significantly facilitate metasurface on-chip or on-package integration with QCL sources. Thanks to the recent advances in QCLs technology [16], we envision the realization of programmable multifunctional THz emitters operating at room temperature, opening a wide range of opportunities for a novel class of research-oriented and industrial-oriented THz products.

## 4. Non-linear Optics

Non-linear optics examines the light behavior in nonlinear media [216]. Although the nonlinear optical responses of conventional materials are weak, nonlinear optical materials usually play an important role in photonics (i.e., photon generation, manipulation, transmission, detection, and imaging [216]). Currently, there is an active search for suitable THz nonlinear material systems with efficient conversion and a small material footprint. Ideally, the material system should allow for on-chip integration and room-temperature operation. In this regard, 2D materials are highly promising [217], particularly after the recent demonstration of exceptionally large nonlinear THz susceptibility in graphene [28].

The nonlinear interaction of electrons in graphene with intense THz fields, originating from both interband and intraband electron dynamics, has long been predicted and debated by a number of papers, built around the existence of a coherent electronic response to a driving THz field [218-220]. However, only recently, thanks to some pioneering experimental works, the understanding in the field of nonlinear THz optics with graphene is experiencing a relevant expansion both in the domains of saturable absorption (SA) [221,222] and high-harmonic generation (HHG) [28,223]. Basically, THz nonlinearity is linked to the interplay between THz response and the collective thermodynamics of carriers in graphene: it is a thermal, noncoherent mechanism. One of the essential consequences of the Dirac-type electronic band structure of graphene is the extremely low heat capacitance of the electronic subsystem. As discussed earlier, this feature allows for the onset of large thermal gradients in the electronic distribution under illumination as a result of heat accumulation. The occurrence of this effect is broadband: the electronic population can be efficiently heated both by interband or intraband absorption processes [27]. In the case of THz illumination, the electron-optical phonons cooling channel is energetically inhibited [27], giving rise  an even more pronounced hot-carrier distribution, which corresponds to a lower conductivity with respect to the cold state. Therefore, the ultrafast carrier thermodynamics in graphene is tightly bound to the electrodynamics and, ultimately, to the ultrafast reshaping of the material optical response. An outstanding consequence of this connection and of the ~ps cooling timescales of hot-carriers is the strong nonlinear response to electric fields in the THz range, which persists at room temperature and under ambient exposure.

As a result, graphene is a highly efficient THz nonlinear absorber [224], with a power transmission modulation of about 50% per single monolayer [223,225], and an efficient frequency multiplier and, thanks to nonlinear coefficients ($X^{(3)} \sim 10^{-9} m^2 V^{-2}$) that are many orders of magnitude larger with respect to those of all other known THz materials [223], allows for the generation of multiple THz harmonics, up to the ninth order [226]. Interestingly, recent findings [227] have demonstrated that this strong THz nonlinearity can be tuned by changing the carrier density (the Fermi level) in graphene via electrostatic gating. This offers a simple route to control both SA and HHG phenomena, opening up a new realm of THz nonlinear optical components.

## 4.1 Saturable absorption

Saturable absorption is a non-parametric third-order nonlinear optical process and corresponds to an intensity-dependent change of the refractive index, due to the field-dependent nonlinear optical conductivity of a material. Owing to the ease of integration in photonic devices, such as fibers, waveguides and resonators, 2D materials have been widely used to realize SAs in the visible to telecom range [228-232]. However, so far, only graphene has been demonstrated to operate as a saturable absorber in the THz range [221,222,225]. In this frequency range, the absorption in doped graphene ($E_F > k_B T$) is dominated by the intraband (Drude) conductivity, which changes upon absorption as a consequence of carrier heating. This effect has been extensively studied both theoretically and experimentally and can be microscopically understood as originating from a reduced screening of the Coulomb interaction of the heated carrier distribution with impurities: carriers with higher energy (temperature) have a lower mobility, and therefore a lower conductivity [233]. An alternative compelling explanation for the reduction of the intraband (THz) conductivity is based on spectral weight conservation [225]. The increased electronic temperature leads to a downshift of the chemical potential, which, in turn, translates into an increased spectral weight for interband transitions that is accompanied by a corresponding decreased weight for Drude absorption: a reduced low-frequency conductivity [56,223]. This THz-induced negative photoconductivity then translates in a reduction, i.e. saturation or bleaching, of absorption.

THz graphene based saturable absorbers (GSAs) have been realized in chemical vapor deposition (CVD)-grown [221], epitaxial [222] and ink-jet printed graphene with modulation depths up to 80% under pump intensity of ~5 Wcm$^{-2}$ [234] (Figure 5a) and with the possibility of tuning via electrostatic gating [227]. Interestingly, the flexibility and versatility of large-area graphene thin films open a wide range of opportunities for the integration of low-footprint nonlinear components with pre-existing photonic devices.

On-chip solution-processed multilayer graphene saturable absorber (GSA) reflectors recently proved to be a key non-linear component for engineering record dynamic range THz quantum cascade laser frequency combs based on a monolithic coupled-cavity architecture. Frequency and phase locking of the modes, prerequisite of comb formation, is here obtained through four-wave-mixing, which is driven by either fast saturable gain or loss in the QCL cavity [235]. The former mechanism leads to a frequency modulated output, while the latter is associated with amplitude modulation. In a THz QCL frequency comb, both frequency and amplitude modulations are typically present [236,237] and act simultaneously. In the engineered coupled cavity architecture (Figure 5b) [238], the gain and absorption are spatially separated, so they do not average out to a local net gain/loss, and create a spatially dependent profile within the cavity. As a result of the interaction of the field emitted by the QCL with the inherently fast GSA [234], and the related reinjection of this field into the laser cavity, the fast saturable loss of the GSA contributes to the locking between the modes, which manifests itself through the observation of extremely narrowed beatnotes. In addition to the saturable absorption effect inside graphene, the reflector also features Fresnel reflection on the graphene surface. Since this component is spatially separated from the saturable absorption occurring inside graphene, it might contribute to frequency comb stabilization through the same mechanism as the fast saturable gain in the active region. Given its ultrafast saturation dynamics [26], on the same timescales as the gain recovery time in a THz QCL, graphene is regarded as one of the most promising candidates for the realization of the long-sought passive mode-locking in such a class of optoelectronic devices

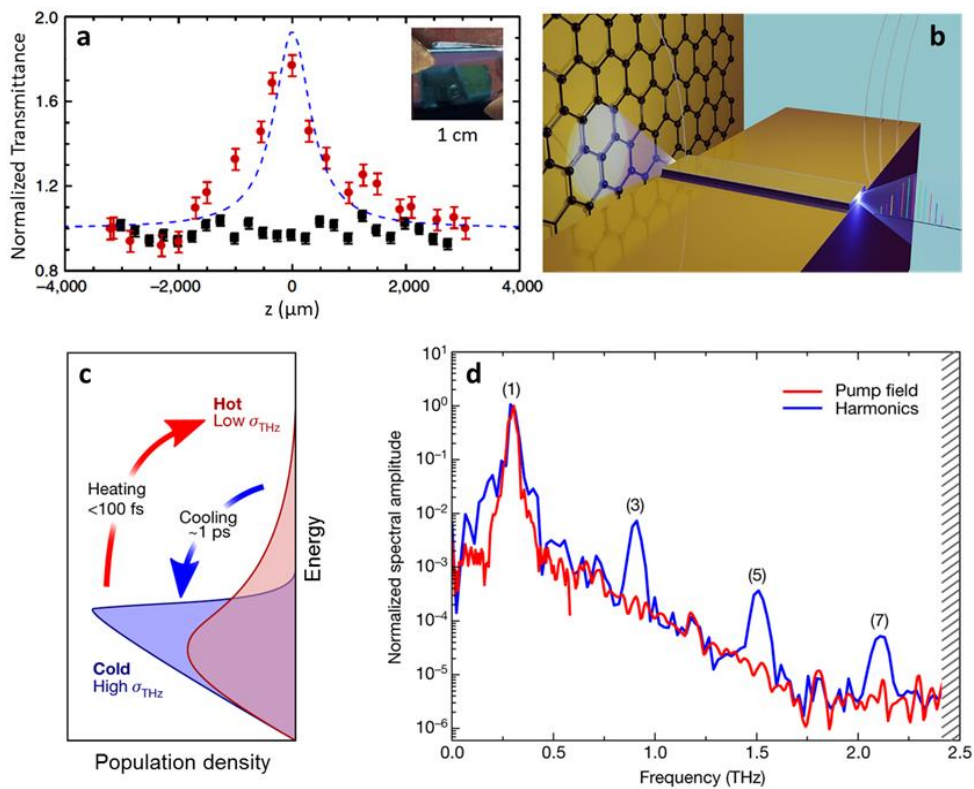

**Figure 5.** (a) Modulation depth of a large-area GSA realized from liquid phase exfoliation of graphite, measured with a *z*-scan technique [234]; the inset shows a photograph of the GSA. (b) Solution-processed GSAs can be integrated with miniaturized THz QCL frequency combs to improve the phase coherence between lasing modes, enabling high-power and stable frequency comb operation [238]. (c) The heating and cooling processes in graphene occur on different timescales. The sheet optical conductivity $\sigma_{THz}$ (the absorption) strongly depends on the carrier temperature. Under THz excitation (period ~ps) $\sigma_{THz}$ is modulated and changes during the THz cycle, leading to a strong non-linear response. (d) THz high-harmonic generation from a CVD-grown single layer graphene (blue line). The red line shows the amplitude spectrum of the incident pump THz wave [28]. Figure (a) is adapted from V. Bianchi *et al.* Nat. Commun. 8, 15763 (2017). Figure (b) is reprinted with permission from F.P. Mezzapesa *et al.* ACS Photon. 7, 3489 (2020). Copyright 2020, American Chemical Society. Figures (c) and (d) are reprinted with permission from H.A. Hafez *et al.* Nature 561, 507 (2018). Copyright 2018, Springer Nature.

## 4.2 High Harmonic generation

Owing to the collective thermodynamic response of Dirac carriers to a THz excitation, graphene presents giant odd-order THz susceptibilities if compared to conventional nonlinear crystals [239] and it has been argued that it is the most nonlinear material known to date [223]. This property is strictly related to the ultrafast carrier heating and cooling processes in the atomically thin graphene layer. In fact, the intraband conductivity depends on the carrier temperature, which can be modified by an impinging THz pulse. The conversion of THz energy into electronic heat is quasi-instantaneous (~50 fs) and the time scale over which the carrier population remains in the hot-state is set by the ~ps cooling time (Figure 5c) [56]. If the variation of $T_e$ takes place on a timescale comparable to a cycle of the THz field, the real part of the complex conductivity varies during a single cycle, leading to a significant temporal distortion of the THz field, which results in the generation of higher-order harmonics. This effect has been used on large-area CVD graphene to excite third-, fifth- and seventh-order harmonics starting from a multi-cycle pulse centered at 0.3 THz with peak field of ~10 kVcm⁻¹ (Figure 5d) [28]. Interestingly, by exploiting the carrier density tunability via electrostatic gating, HHG can be tuned and modulated [227]. Moreover, the use of planar metallic structures can increase the nonlinear light-matter interaction by locally increasing the field strength, reaching $X^{(3)}$~3×10⁻⁸m²V⁻² with the aid of metallic gratings [226].

## 4.3 Challenges and opportunities

The recently demonstrated outstanding nonlinear properties of graphene [28,225] open prospects for the generation and modulation of light in small footprint architectures, taking advantage of the easy integration and room temperature operation, therefore ensuring low-power consumption.

The integration of GSA with QCL frequency combs [238] can open new prospects for the generation of ultra-short THz pulses with compact and potentially portable systems, given the recent advances in high temperature QCLs [16].

Furthermore, a possible disruptive application of HHG is represented by all-electronic frequency mixing and upconversion of sub-THz signals generated by CMOS platforms, with the potential of outcompete the current generation of diode-based ultrahigh frequency mixers [240] in terms of conversion efficiency, operation bandwidth and chip cost, promising a great impact in THz information and communication technologies.

Eventually, extensive research has been also recently devoted to investigate non-linear effects in graphene related 2D materials [241]. For example, BP displays a strong mid-IR saturable absorption, driven by carrier dynamics faster than those observed in graphene [242], however, no demonstration of nonlinear THz response has been observed to date, except for graphene.

## 5. Terahertz Nanoscience

Infrared nanoscience with two-dimensional nanomaterials has been a particularly vibrant research field, in the last few years. Examples include ultrastrong light-matter interactions [243,244], hydrodynamic effects [245-247], single-plasmon nonlinearities [200,248], polaritonic quantization [249-251] or topological nanophotonics [252,253].

In addition to these intrinsic quantum phenomena, 2D material systems can also be used as sensitive probes for inspecting the quantum properties of the material that carries the nanophotonic modes. Systems based on 2D nanomaterials indeed provide an intriguing platform for capturing plasmons, plasmon-polaritons [254] — hybrid light-matter modes involving the collective oscillations of mobile charges, or phonon-polaritons [255], whose wavelength can be electrically controlled, through the exploitation of their confined electronic systems [190,191,256-258]. The possibility of hybridizing collective electronic motion with light in so-called surface polaritons has indeed made these materials a versatile platform for extreme light confinement and tailored nanophotonics.

Furthermore, 2D materials are fully compatible with a wide range of substrates including flexible and transparent ones, meaning that, if placed on chip with flat integrated optical circuits, they can allow maximal interaction with light, therefore optimally utilizing their novel and versatile properties for a wealth of applications in transformation optics [259] and high-speed optical communications [260-262].

To properly control light at the nanoscale, the impinging optical field needs to be compressed into small volumes, well below the diffraction limit. This is even more difficult at THz frequencies that have a very long wavelength and below the nanometer scale, due to the increased momentum mismatch between the far-field photon and the excited optical mode [253]. In this context, 2D materials can enable pushing light-matter interaction into the atomic limit [263,264], owing to their unique optical response. Therefore, capturing the aforementioned modes in 2D materials, at THz frequencies, requires sophisticated optical techniques capable of imaging at their exact frequency or mapping their long-range propagation, usually with sub-micrometer resolution.

In the visible and the infrared range, sub-10-nm resolution is achieved by scattering near-field optical microscopy (s-SNOM) [265], where an atomic force microscope (AFM) tip converts the incident light into strongly concentrated fields at the tip apex (nanofocus) to locally excite molecular vibrations, plasmons or phonons in the sample [266,267]. The spatial resolution is thus determined by the tip apex size rather than the radiation wavelength, but limited by the weak scattering efficiency of the tip. Indeed, in the THz range, the scattering efficiency of AFM tips is prohibitively low [268,269], demanding the use of powerful gas lasers combined with cooled bolometers, or THz time-domain spectroscopy (THz-TDS) systems, which can detect very weak scattered fields, but with limited spectral resolution and slow image acquisition [270]. Recently, detector-less s-SNOM has been proposed as an innovative tool to reach nm resolution in a compact configuration (Figure 6a) [257,271].

At the other end of the spectrum of length scales targeted by THz microscopy are applications associated with components for emerging THz communication, where the challenge is to map THz absorption properties and THz field distributions in large-area samples. The s-SNOM approach is not appropriate for these applications, due to AFM instabilities arising during scans of rough or soft surfaces at scan speeds required for large-area imaging. As such, THz microscopy approaches that can address these applications are based on near-field aperture probes with integrated sub-wavelength size THz detectors (a-SNOM) [272-274]. Conventional a-SNOM architectures are fundamentally limited in spatial resolution by the active region size of the employed THz electro-optical detectors, and the state-of-the-art resolution has stagnated at the level of 2–10 μm [275,276]. Recently, the first THz a-SNOM probe comprising a sensitive THz nano-detector exploiting an active 2D nanomaterial (encapsulated BP flake) embedded within the aperture has been demonstrated (Figure 6b) [277]. It enables room-temperature THz a-SNOM imaging and spectroscopy systems operating with spatial resolution of ~$\lambda$/100 and with an extremely high sensitivity controlled by the thermoelectric detection dynamics engineered in the probe active element [274,277].

## 5.1 Challenge and opportunities

THz nanoscopy is an appealing tool for several application fields. In nanoscale physics and electronics it can allow mapping THz plasmons in 2D nanomaterials (e.g. graphene [205,278,279], topological insulators [280], phosphorene, silicene, and their heterostructures), which can reveal groundbreaking perspectives for the development of a completely novel class of ultra-sensitive plasmonic devices and for ultrafast components like saturable absorbers and modulators.

Local electronic properties of vdW heterostructures and THz surface plasmons can also be simultaneously mapped on different scales, with two complementary methods for plasmon mapping: (*i*) using nanofocused excitation by the s-SNOM probe and (*ii*) using plasmon excitation within resonant periodic structures (e.g. micro-ribbon arrays) [281].

THz nanoscopy can also allow the investigation of electronic states (energies and spatial distribution) of dopants in quantum dots, quantum wires [271], and nanostructured interfaces between dissimilar materials [282]. This will allow understanding transport phenomena in low-dimensionality optoelectronics with an unprecedented spatial and spectral control, opening the path to the design of high efficiency THz modulators, switches, polarizers, spin-manipulator systems, single-photon detectors. This will represent a major step towards the possibility of probing local quantum states, underpinning next generation quantum technologies.

Finally, 2D nanomaterials can be also innovatively employed to engineer ultrafast switches for electronic waves [258]. By tracing in time the expansion of the plasmon waves of a black-phosphorus based heterostructure with extreme slow motion snapshots, the experimental on/off switching of waves on the electron sea was demonstrated, laying the foundation for future plasma-electronics. Such a sophisticated experiment proved that the plasmon switching times are on the femtosecond scale, many orders of magnitude faster than the fastest existing transistors [283], potentially allowing to speed up future electronics many times over.

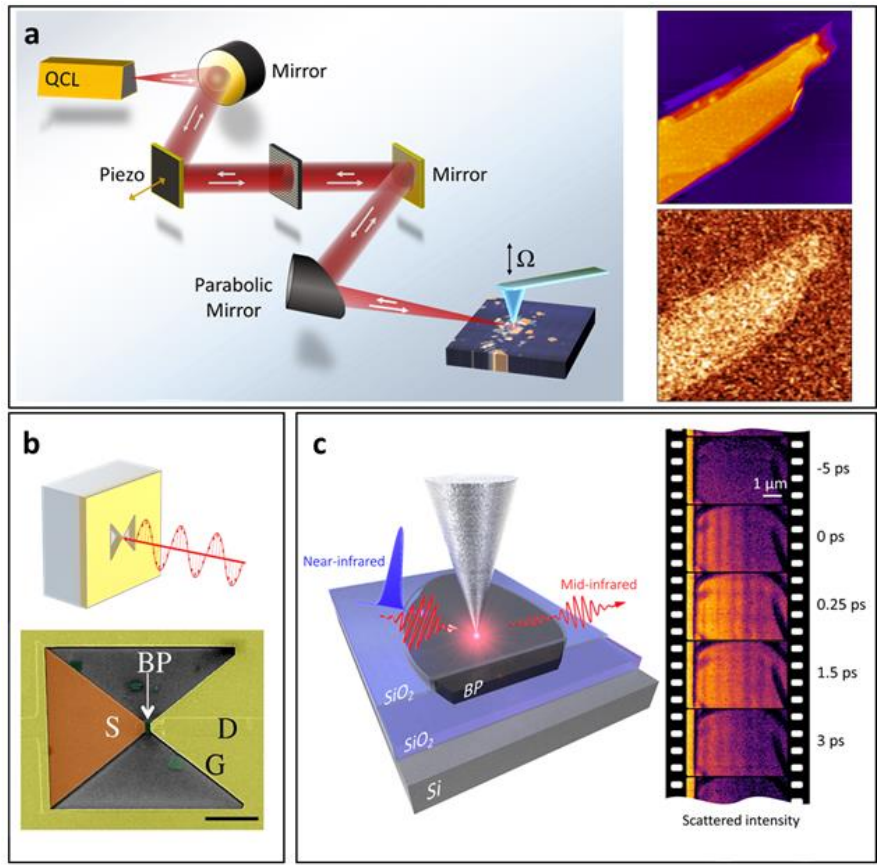

**Figure 6.** (a) Schematic representation of a self-detection scattering type near-field optical microscope based on a THz QCL, which simultaneously operates as source and *coherent* detector of radiation by means of the self-mixing effect [257]. The insets show the topographic map of a black-phosphorus flake and the corresponding scattered near-field signal as retrieved by the self-detection scheme. (b) Schematic view and false color scanning electron microscope (SEM) image of an aperture-based near-field probe with an embedded BP-based THz nanodetector [277]. (c) THz SNOM can be employed in the excitation and detection of plasmon polaritons—hybrid light matter modes involving the collective oscillation of mobile charges, allowing to trace these modes in time, energy and space. Recent pump-probe experiments on two-dimensional black phosphorus revealed the ultrafast ignition of propagating polariton modes on femtosecond timescales [258]. Figure (a) is adapted from M.C. Giordano *et al.* Optics Express 26, 18423 (2018). Copyright 2018, Optical Publishing Group. Figure (b) is adapted with permission from O. Mitrofanov *et al.* Sci. Rep. 7, 44240 (2017). Copyright 2017, the Author(s). Figure (c) is adapted with permission from M.A. Huber *et al.* Nat. Nanotechnol. 12, 207 (2017). Copyright 2016, Springer Nature.

## 6. Conclusion

In this perspective article, we discuss the recent advances on 2D-material-based devices that are promising to play a disruptive role in terahertz photonics and optoelectronics. We focus on four major branches: THz detection, modulation, non-linear light-matter interaction and THz nanoscience. In each of these domains, two-dimensional materials present distinctive properties, such as tunability, extreme field confinement, ultrafast dynamics and long-lived collective excitations that can raise them as game-changers for devising novel architectures and designing future groundbreaking technologies.

Such a vibrant research field is still in its early stages of development and the needed technological leap, driven by the promising level of performances, will have to go through chip-scalability, wafer-scale integration, reproducibility and cost-effectiveness. This will also require physicists, chemists and engineers in the community to redirect their focus from the fascinating and vibrant fundamental physics that characterizes low-dimensional materials towards more application-oriented goals.

## Acknowledgements

This work was partly supported by the European Research Council through the Consolidator Grant SPRINT (681379), and by the European Union through the Graphene Flasghip (Core 3), and the Marie Curie H2020-MSCA-ITN-2017, TeraApps (765426).

## Data Availability Statement

The data that support the findings of this study are available from the corresponding author upon reasonable request.